\documentclass[
preprint,
showpacs,
 amsmath,amssymb,
 aps,
prb,
]{revtex4-1}

\usepackage{graphicx}
\usepackage{dcolumn}
\usepackage{bm}
\usepackage{hyperref}

\begin{document}

\title{Thermoelectric power of Ba(Fe$_{1-x}$Ru$_x$)$_2$As$_2$ and Ba(Fe$_{1-x}$Co$_x$)$_2$As$_2$: possible changes of Fermi surface with and without changes in electron count }

\author{H. Hodovanets}
\author{E. D. Mun}
\author{A. Thaler}
\author{S. L. Bud'ko}
\author{P. C. Canfield}
\affiliation{Ames Laboratory, US DOE and Department of Physics and Astronomy, Iowa State University, Ames, Iowa 50011, USA}

\date{\today}

\begin{abstract}
Temperature-dependent, in-plane, thermoelectric power (TEP) data are presented for Ba(Fe$_{1-x}$Ru$_x$)$_2$As$_2$ ($0 \leq x \leq 0.36$)  single crystals. The previously outlined $x - T$ phase diagram for this system is confirmed. The analysis of TEP evolution with Ru-doping suggests significant changes in the electronic structure, correlations and/or scattering occurring near $\sim 7\%$ and $\sim 30\%$ of Ru-doping levels. These results are compared with an extended set of TEP data for the electron-doped  Ba(Fe$_{1-x}$Co$_x$)$_2$As$_2$  series.
 
\end{abstract}


\maketitle

The recent discovery of families of Fe-As containing materials supporting superconductivity with elevated transition temperatures, $T_c$,  has attracted the attention of the condensed matter physics community.  \cite{japxxa,njp09a,phy09a,sus10a} From the very beginning, details of the electronic structure of these materials were considered to be  of importance for magnetism and superconductivity, \cite{sad08a,iva08a,izy08a,liu09a,maz09a,col10a,nin08b,tor08a,gol08a,kre08a,gol09a,liu09b,can09a} since in most cases superconductivity was achieved by doping or application of pressure. At least in the particular case of electron-doping of the BaFe$_2$As$_2$  with a transition metal, it is thought that for superconductivity to appear, the structural/magnetic transition temperature should be suppressed enough {\it and} the additional electron count caused by doping should be within the certain window. \cite{nin09a,can09b,can10a}    For the case of the most intensely studied, Ba(Fe$_{1-x}$Co$_x$)$_2$As$_2$ family, \cite{sef08a,nin08a,chu09a} the onset of superconductivity was shown \cite{mun09a,liu10a} to coincide with a Lifshitz transition \cite{lif60a} [change of a Fermi surface (FS) topology]. 

Whereas angle-resolved photoemission spectroscopy (ARPES) or quantum oscillations are extremely important in giving a detailed description of the FS evolution through a Lifshitz transition, in many cases, less demanding, transport measurements, in particular thermoelectric power, were proven to be very sensitive to the existence of Lifshitz transitions. \cite{var89a,bla94a} Indeed, Hall effect and, more notably, TEP displayed a clear anomaly at the low-doping Lifshitz transition in  the Ba(Fe$_{1-x}$Co$_x$)$_2$As$_2$ and the  Ba(Fe$_{1-x}$Cu$_x$)$_2$As$_2$  series. \cite{mun09a} Additionally, in some range of Co-concentrations, very large, negative ($\approx -50$ $\mu$V/K) TEP values, often associated with a presence of strong electronic correlations, were observed.

Inasmuch as the simple concept of a rigid band appears to give some qualitative understanding of the evolution of physical properties with electron doping,  there is less understanding of the salient parameters governing the evolution of the physical properties under pressure or with isoelectronic doping.  One of the recent examples of the latter is Ru-substitution for Fe in BaFe$_2$As$_2$. Studies by several groups, using single crystals \cite{rul10a,tha10a} as well as polycrystalline \cite{sha10a} samples of  Ba(Fe$_{1-x}$Ru$_x$)$_2$As$_2$, suggest that the structural/magnetic transition is fully suppressed for $ x \approx 0.30$, and a superconducting dome with the maximum $T_c$ of $\sim 20$ K is observed for $0.2 \lesssim x \lesssim 0.4$. An initial band-structural study of Ru substitution in BaFe$_2$As$_2$ \cite{zha09a}  suggested that the qualitative difference between Fermi surfaces of pure BaFe$_2$As$_2$ and pure BaRu$_2$As$_2$ is the existence of three-dimensional, closed, hole pockets centered near $Z$-point in the latter, in contrast to open, corrugated, hole cylinders along $\Gamma - Z$ in the former. An ARPES study of Ba(Fe$_{0.65}$Ru$_{0.35}$)$_2$As$_2$ resolved several FS pockets but presented no evidence of any topological changes from the parent BaFe$_2$As$_2$ compound. \cite{bro10a}

In this work we present measurements of the in-plane, temperature-dependent, TEP on  Ba(Fe$_{1-x}$Ru$_x$)$_2$As$_2$ ($0 \leq x \leq 0.36$)  single crystals, with the goals of confirming and refining $x - T$ phase diagram, and  determining  Ru-concentration ranges where significant FS changes may possibly occur. We compare the data for the Ba(Fe$_{1-x}$Ru$_x$)$_2$As$_2$ series with the TEP results for the Ba(Fe$_{1-x}$Co$_x$)$_2$As$_2$, by analysis of the data reported in Ref. \onlinecite{mun09a} in addition to new data taken for extended (primarily higher) range of Co-concentrations.
\\

Single crystals of  Ba(Fe$_{1-x}$Ru$_x$)$_2$As$_2$ ($0 \leq x \leq 0.36$)  and  Ba(Fe$_{1-x}$Co$_x$)$_2$As$_2$  ($0 \leq x \leq 0.42$) were grown out of self flux using conventional high temperature solution growth techniques, as described in detail in Refs. \onlinecite{nin08a,tha10a}. Elemental analysis of the single crystals was performed using wavelength dispersive x-ray spectroscopy (WDS) in a JEOL JXA-8200 electron microprobe. Measured (as opposed to nominal) Ru- and Co-concentrations are used in the text. Physical properties of the majority of the samples in this study were described in detail in previous publications. \cite{nin08a,tha10a} TEP measurements were carried out by a dc, alternating current (two heaters and two thermometers) technique \cite{mun10a} over the temperature range between 2 K and 300 K using a Quantum Design PPMS to provide the temperature environment.
\\

\begin{figure}
\begin{center}
\includegraphics[angle=0,width=120mm]{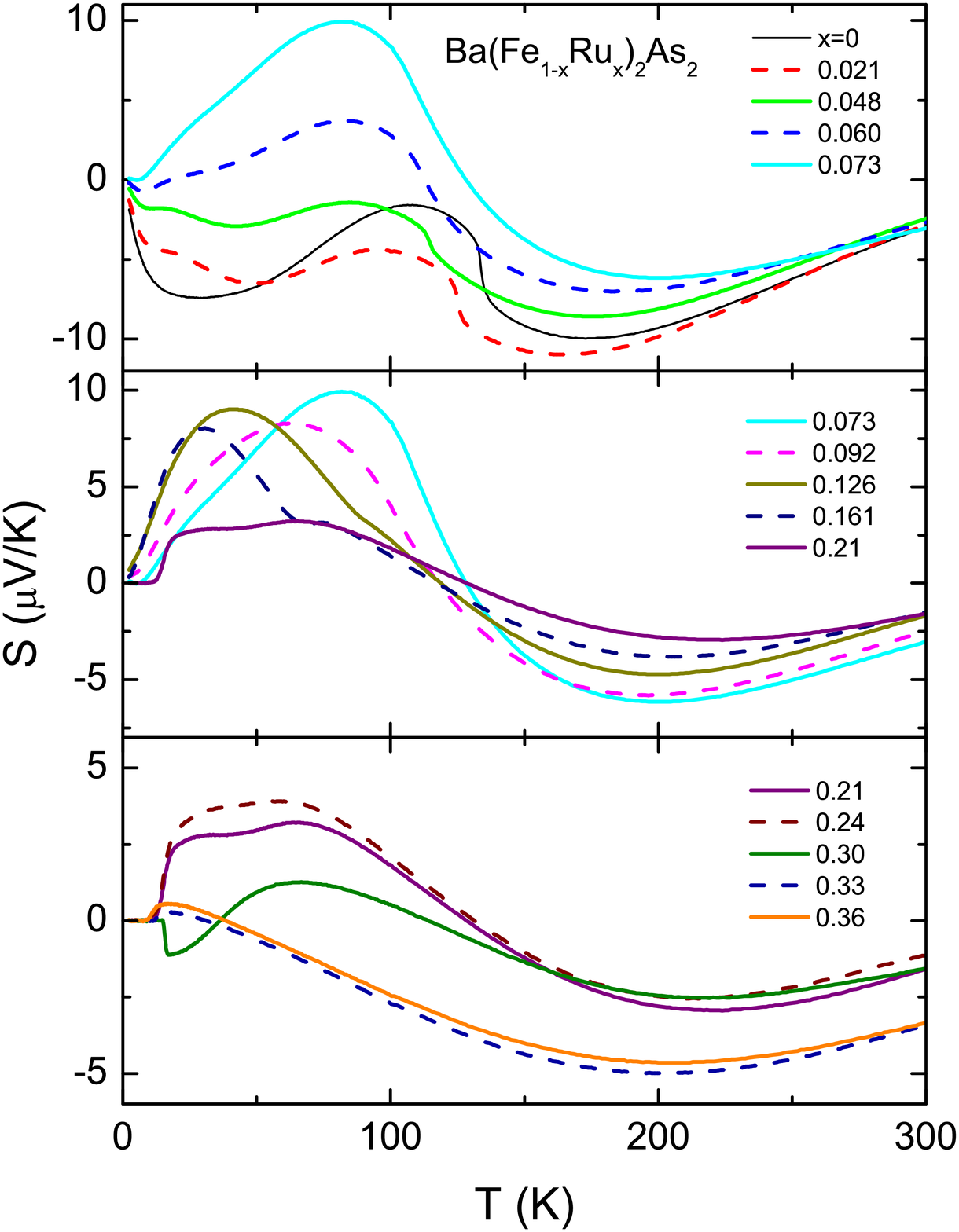}
\end{center}
\caption{\label{F1} (Color online) In-plane TEP of the  Ba(Fe$_{1-x}$Ru$_x$)$_2$As$_2$ ($0 \leq x \leq 0.36$) single crystals. The plot is divided in three panels for clarity. The curves for $x = 0.073$ and $x = 0.21$ are repeated on two panels each for continuity.}
\end{figure}

\begin{figure}
\begin{center}
\includegraphics[angle=0,width=120mm]{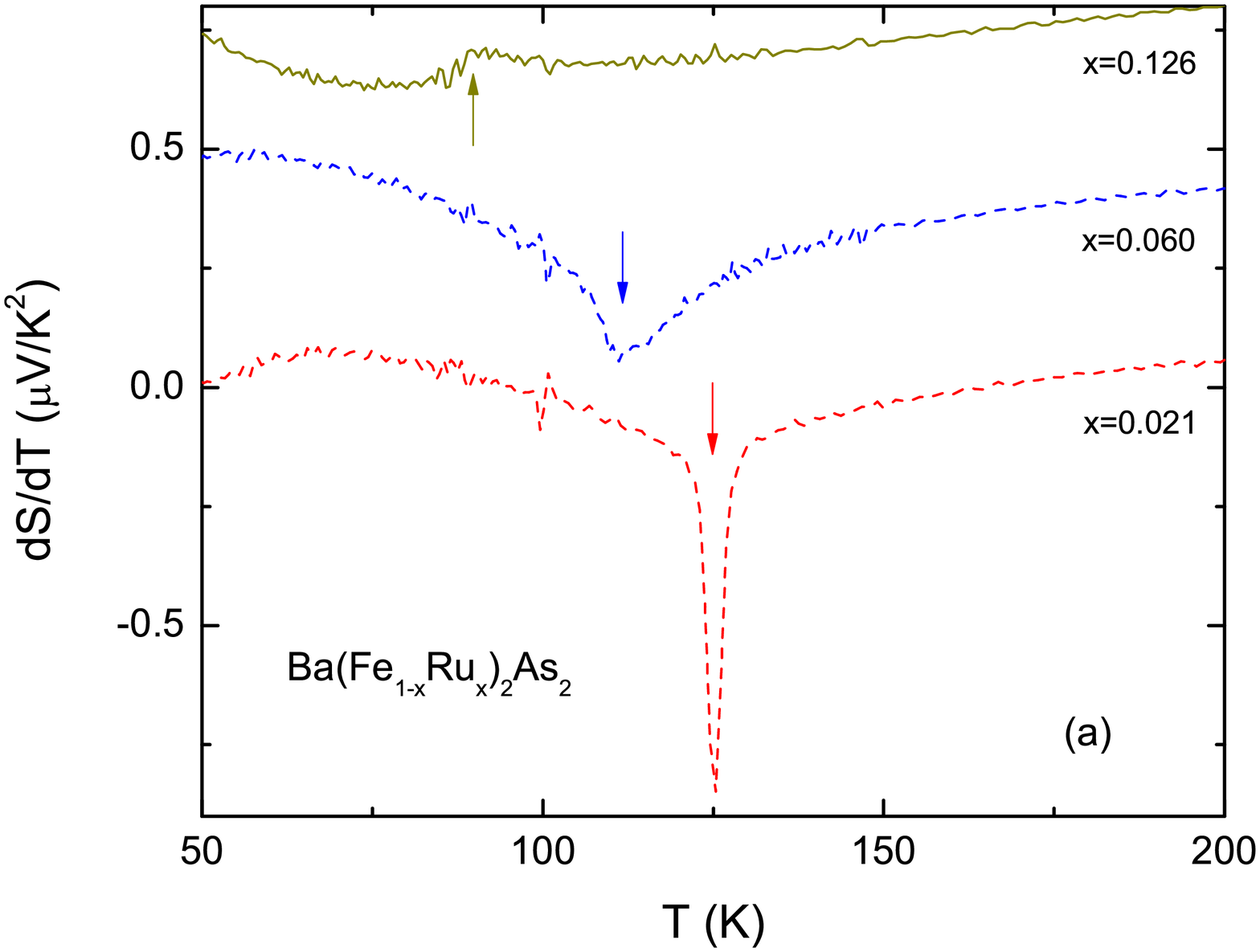}
\includegraphics[angle=0,width=120mm]{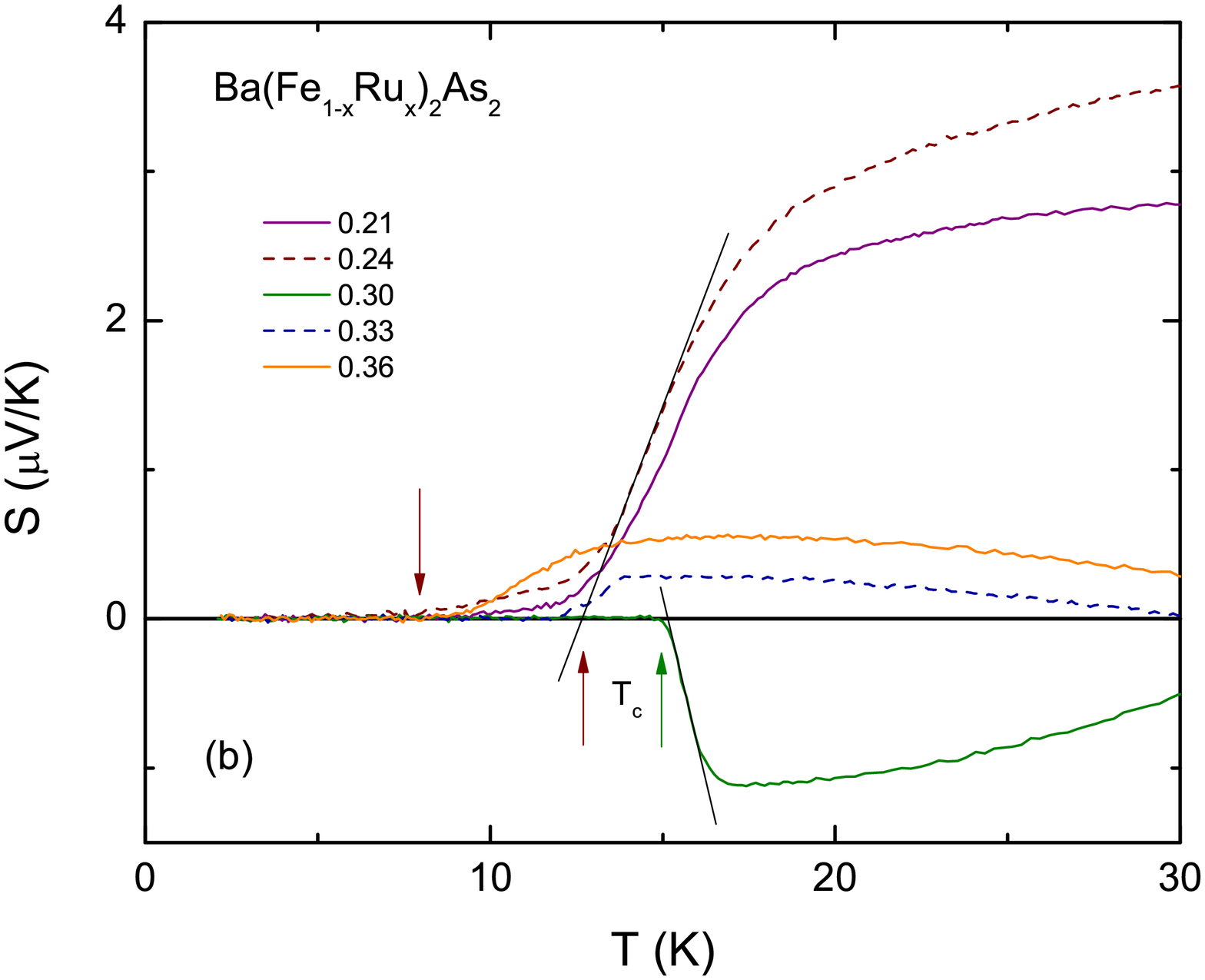}
\end{center}
\caption{\label{F2} (Color online) (a) Derivatives, $dS/dT$, for representative concentrations, with the structural/magnetic transition temperatures marked with arrows. The data for $x = 0.060$ and $x = 0.126$ are shifted along the y-axis by 0.4 and 0.8 $\mu$V/K$^2$ for clarity. (b) Low temperature $S(T)$ curves for $0.21 \leq x \leq 0.36$ with the $T_c$ criteria marked with arrows (see text for details). }
\end{figure}

\begin{figure}
\begin{center}
\includegraphics[angle=0,width=120mm]{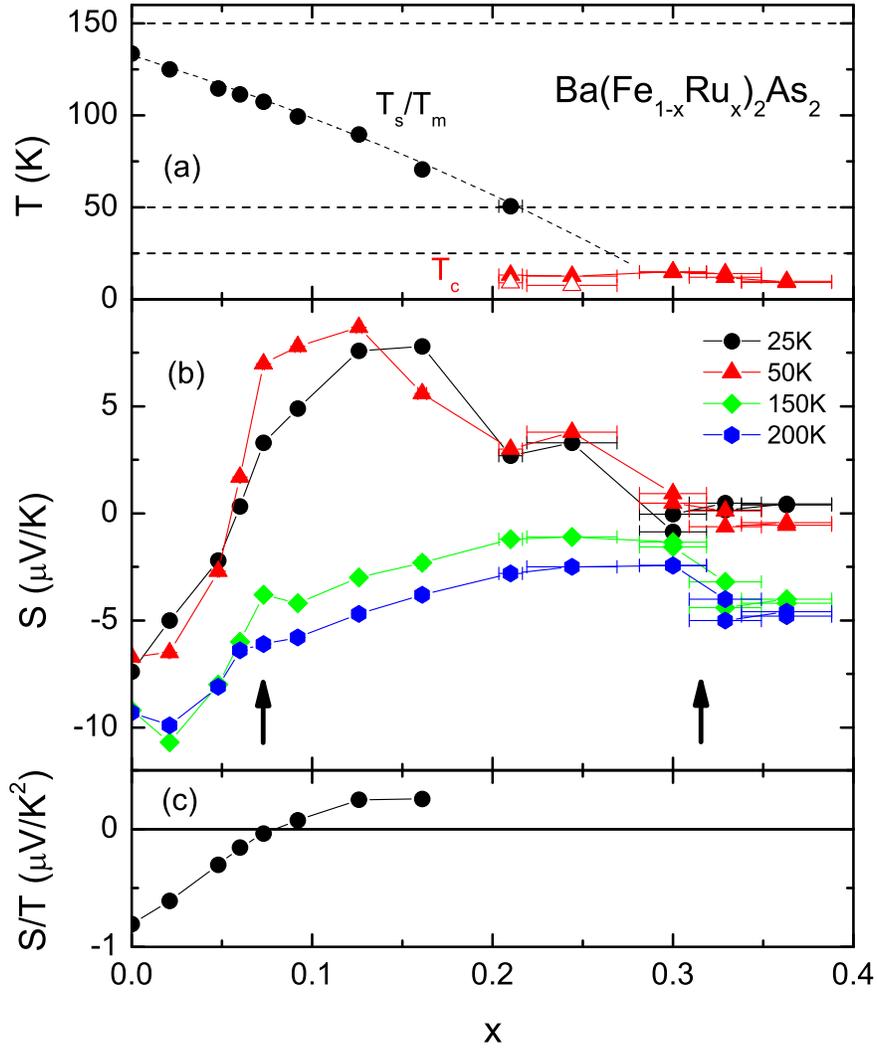}
\end{center}
\caption{\label{F3} (Color online) (a) $x - T$ phase diagram for  Ba(Fe$_{1-x}$Ru$_x$)$_2$As$_2$ obtained from the TEP measurements. $T_s/T_m$ denotes the structural/magnetic transition, $T_c$ - superconducting transition, lines through the experimental points are guides for the eye, the criteria used are explained in the text, with the open triangles showing $T_c$ from $S = 0$ criterion. The horizontal lines correspond to 25, 50  and 150 K. (b) $x$-dependent TEP of Ba(Fe$_{1-x}$Ru$_x$)$_2$As$_2$ at fixed, 25, 50, 150, and 200 K temperatures. Arrows mark the regions of anomalous $S(x)|_{T=const}$ behavior. (c) Low-temperature values of $S/T$ for non-superconducting  Ba(Fe$_{1-x}$Ru$_x$)$_2$As$_2$ samples. Error bars correspond to $\pm \sigma$ (standard deviation) in Ru concentration as determined by WDS.}
\end{figure}

Temperature-dependent, in-plane, TEP data for Ba(Fe$_{1-x}$Ru$_x$)$_2$As$_2$ ($0 \leq x \leq 0.36$) single crystals are shown in Fig. \ref{F1}. In contrast to Ba(Fe$_{1-x}$Co$_x$)$_2$As$_2$ (Ref. \onlinecite{mun09a}) the absolute values of TEP do not exceed $\sim 10$ $\mu$V/K. For all concentrations measured, a broad minimum is observed in the 150-200 K temperature range. In addition, multiple, broad, features (the origins of which are unclear at this point) are observed for many Ru-concentrations. For $0.21 \leq x \leq 0.36$, zero TEP at low temperatures, corresponding to the superconducting state, is clearly seen in the data. The criteria used for constructing of a $x - T$ phase diagram from the TEP data are shown in Fig. \ref{F2}. For the structural/magnetic transition an extremum in the derivative, $dS/dT$, is used to infer a critical temperature. As already noted from resistance and susceptibility data, \cite{tha10a} with Ru doping, the structural/magnetic transition is suppressed, the characteristic feature broadens, but no signature of split transitions is observed. It is noteworthy that, starting from $x = 0.126$, the characteristic feature marking the structural/magnetic transition changes from a local minimum to local maximum (Fig. \ref{F2}a). For superconducting transitions, an offset criterion (as shown for $x = 0.30$ in Fig. \ref{F2}b) was used to infer $T_c$. For two concentrations, $x = 0.21$ and 0.24, $S(T)$ data have a significant shoulders at the superconducting transition. In these two cases two criteria, offset and $S(T) = 0$ were used (marked by arrows for $x = 0.24$ in Fig. \ref{F2}b). The phase diagram obtained from the TEP measurements (Fig. \ref{F3}a) is consistent with that reported in Ref. \onlinecite{tha10a}: the structural/magnetic transition is suppressed by $x \sim 0.3$ and the superconducting dome exists between approximately $x = 0.2$ and $x = 0.4$.

There are several approaches that allow for the detection of changes in the electronic structure from TEP measurements. TEP at fixed temperature plotted as a function of a control parameter (Ru- or Co-concentration in our case) is expected to show anomalous behavior at Lifshitz transitions. \cite{var89a,bla94a} Figure \ref{F3}b shows the doping dependence of TEP at selected, fixed temperatures. For the three highest Ru concentrations, results from measurements on two samples each are shown. The data for $T = 25$ K and 50 K may have an additional feature (between $x \sim 0.2$ and $x \sim 0.3$)  caused by crossing of the structural/magnetic transition line ($T_s/T_m$  in Fig. \ref{F3}a). The data for $T = 150$ K and 200 K, on the other hand,  correspond to the same, tetragonal, structure and absence of the magnetic order. A change in the doping dependence of TEP at selected, fixed temperature, $S(x)|_{T=const}$, data (for all presented temperatures) is clearly seen at $x \sim 0.07$. Another step-like feature at $x \sim 0.3$  is unambiguous in 150 K and 200 K data and is somewhat obscured, possibly by crossing of the $T_s/T_m$ line, for the low temperature data.

\begin{figure}
\begin{center}
\includegraphics[angle=0,width=120mm]{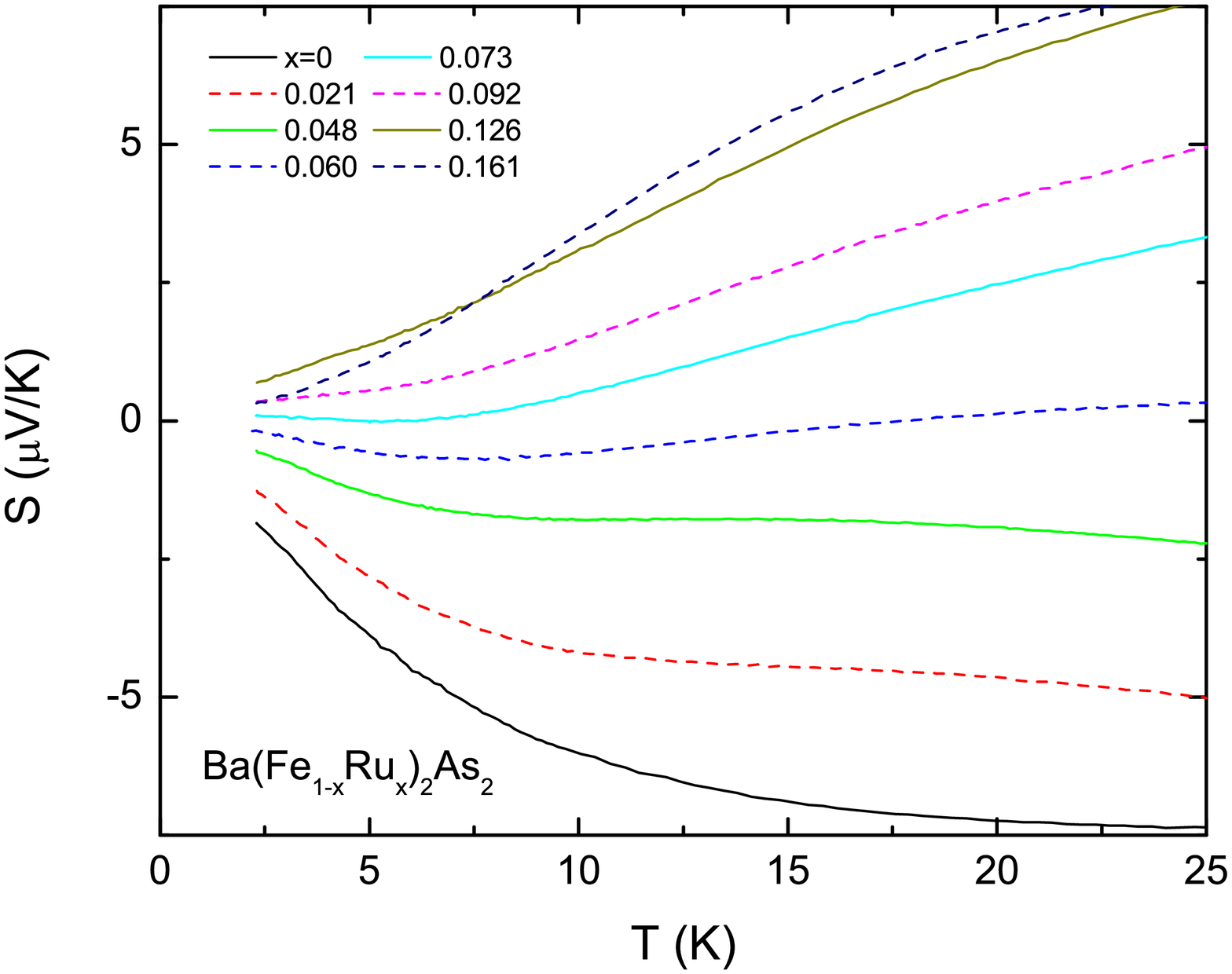}
\end{center}
\caption{\label{F4} (Color online) Low temperature part of in-plane TEP of the  non-superconducting Ba(Fe$_{1-x}$Ru$_x$)$_2$As$_2$ ($0 \leq x \leq 0.161$) single crystals.}
\end{figure}

Another approach relies on the analysis of the low temperature, linear in $T$, coefficient of TEP, $S/T$  (Ref. \onlinecite{beh04a}). For a free electron gas, $S/T$ depends on carrier concentration, density of state and scattering. For real materials the description of TEP becomes very complex. Still, in lieu of comprehensive theory, one can try to look at gross features in $S/T$ as a function of a control parameter. For the non-superconducting Ba(Fe$_{1-x}$Ru$_x$)$_2$As$_2$ samples the low temperature $S/T$ parameter determined from a linear fit of the $S(T)$ data below $\sim 4$ K (see Fig. \ref{F4}), is plotted in Fig. \ref{F3}c. The line crosses zero at $x \approx 0.07$, in the same concentration range where an anomaly in  $S(x)|_{T=const}$ is observed.
\\

\begin{figure}
\begin{center}
\includegraphics[angle=0,width=120mm]{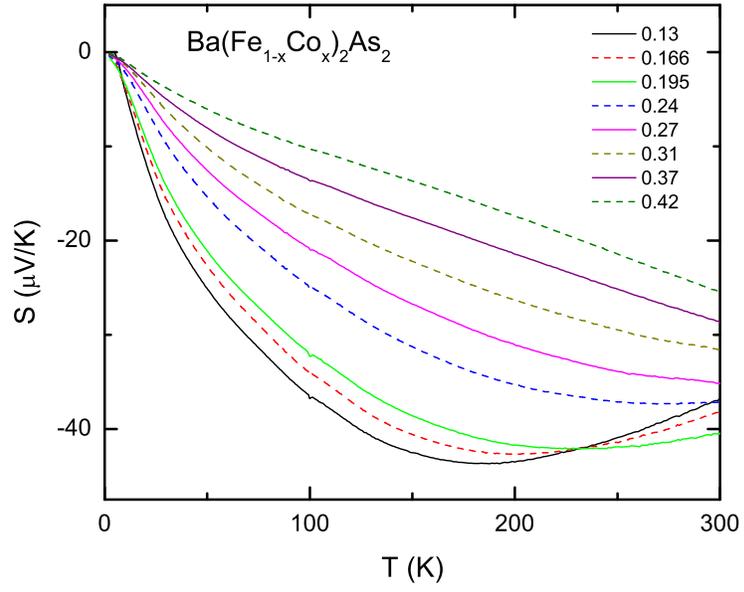}
\end{center}
\caption{\label{F5} (Color online) In-plane TEP of the  Ba(Fe$_{1-x}$Co$_x$)$_2$As$_2$ ($0.13 \leq x \leq 0.42$) single crystals.}
\end{figure}

\begin{figure}
\begin{center}
\includegraphics[angle=0,width=120mm]{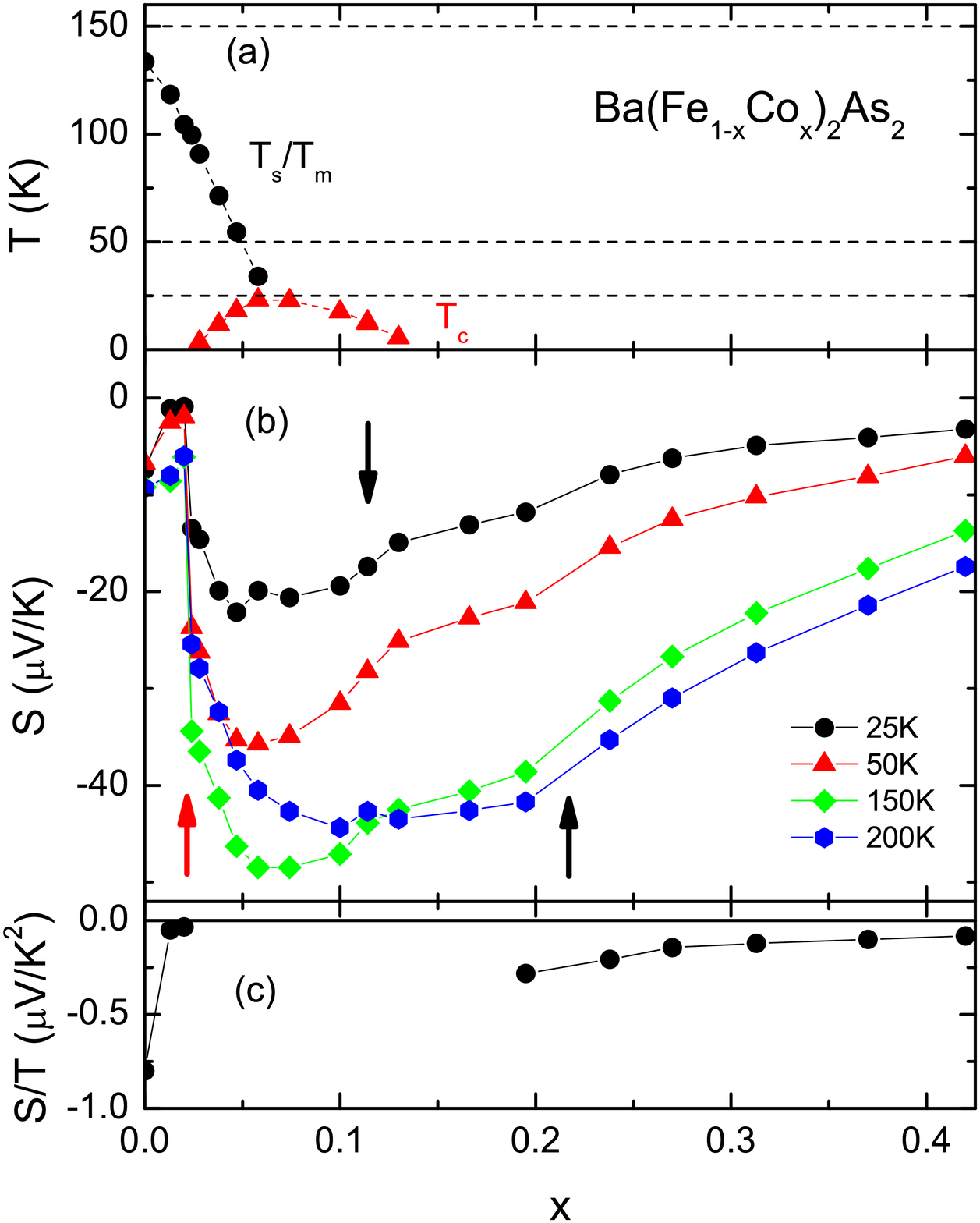}
\end{center}
\caption{\label{F6} (Color online) (a) $x - T$ phase diagram for  Ba(Fe$_{1-x}$Co$_x$)$_2$As$_2$ obtained from the TEP measurements. $T_s/T_m$ denotes the structural/magnetic transitions shown here, schematically as a single line, $T_c$ - superconducting transition, lines through the experimental points are guides for the eye, the criteria used are explained in the text. The horizontal lines correspond to 25, 50 and 150 K. (b) $x$-dependent TEP of Ba(Fe$_{1-x}$Co$_x$)$_2$As$_2$ at fixed, 25, 50, 150, and 200 K temperatures. Arrows mark the regions of anomalous $S(x)|_{T=const}$ behavior. (c) Low-temperature values of $S/T$ for non-superconducting  Ba(Fe$_{1-x}$Co$_x$)$_2$As$_2$ samples.}
\end{figure}

The results above for the  Ba(Fe$_{1-x}$Ru$_x$)$_2$As$_2$ series can be compared with the TEP data for the well-studied electron-doped  Ba(Fe$_{1-x}$Co$_x$)$_2$As$_2$ series. For such comparison the TEP data for $0 \leq x \leq 0.114$ were taken from the previous publication, \cite{mun09a} and new data for $0.13 \leq x \leq 0.42$ (Fig. \ref{F5}), extending far into overdoped, non-superconducting range of Co-concentrations were added. It is noteworthy that in this latter, non-superconducting, range of Co- concentrations the $S(T)$ behavior appears to be qualitatively consistent with that described within a simple two-band 3D model. \cite{sal10a} In the overdoped, non-superconducting range of Co-concentrations the broad local minimum moves up in temperature, out of the measured temperature range for $x > 0.2$, with no detectable sudden change in the $S(T)$  values.  Figure \ref{F6} presents a schematic phase diagram for the  Ba(Fe$_{1-x}$Co$_x$)$_2$As$_2$ series, concentration-dependent TEP for 25 K, 50 K, 150 K, and 200 K, and low temperature $S/T$ values for non-superconducting members of the series. The low-concentration Lifshitz transition, discussed at length in previous publications, \cite{mun09a,liu10a} is clearly seen as an abrupt feature in the $S(x)|_{T = const}$  data and in an approaching to zero low-temperature $S/T$ value. On further Co-doping, two, more subtle, features are observed in the $S(x)|_{T = const}$  data: at $x \sim 0.11$, and at $x \sim 0.22$. The possibility of several Lifshitz transitions in Co-doped BaFe$_2$As$_2$ was suggested in several experimental and band-structural studies, \cite{liu10a,fan09a,bro09a,liu10b} broadly speaking,  in the $x \sim 0.2-0.3$ concentration range.  Our TEP data indicate the concentrations $\sim 0.1$ and $\sim 0.2$  for further, more careful investigation. Two new lines on the $x - T$ phase diagram of the  Ba(Fe$_{1-x}$Co$_x$)$_2$As$_2$ series were suggested based on $c$-axis resistivity measurements and NMR. \cite{tan10a} Rather large slopes of these lines are not consistent with observations based on TEP, future $S(T)$ measurements with $\Delta T \| c$ may be instrumental foe understanding of the origin of these lines. 
\\

From this TEP analysis two ranges of Ru-concentrations, $ x \sim 0.07$ and $\sim 0.3$ are suggested for possible Lifshitz transitions, or other drastic changes in electronic structure, correlations or scattering in Ba(Fe$_{1-x}$Ru$_x$)$_2$As$_2$. Whereas in Ba(Fe$_{1-x}$Co$_x$)$_2$As$_2$, the lower concentration Lifshitz transition coincides with the onset of superconductivity, in the case of Ru-doping there is no obvious feature in the $x - T$ phase diagram at $x \approx 0.07$. Similarly, Hall and TEP anomalies \cite{mun09a}  in Ba(Fe$_{1-x}$Cu$_x$)$_2$As$_2$, that occur at the same extra electron, $e$, value as the lower concentration Lifshitz transition in Ba(Fe$_{1-x}$Co$_x$)$_2$As$_2$, do not signal the occurrence of superconductivity. A change in the Fermi surface topology might be necessary but is not sufficient for superconductivity to occur in BaFe$_2$As$_2$ with doping. \cite{can10a} The second anomaly in TEP of  Ba(Fe$_{1-x}$Ru$_x$)$_2$As$_2$ occurs at the concentration corresponding to complete suppression of the structural/magnetic transition, maximum of the superconducting dome and linear behavior of the normal state resistivity. Even though the physical picture behind the remarkably similar, anomalous, behavior of a number of properties of Fe - As based materials driven to such region of the phase diagram either by different dopings or by pressure is not understood, it is clear that TEP is able to delineate this region as well. For the  Ba(Fe$_{1-x}$Co$_x$)$_2$As$_2$ series, two new anomalies were observed, one in the overdoped half of the superconducting dome on the overdoped side, and another one, in the non-superconducting, overdoped, part of the phase diagram,  beyond the dome. 
\\

To summarize, the temperature-dependent in-plane TEP in Ba(Fe$_{1-x}$Ru$_x$)$_2$As$_2$ ($0 \leq x \leq 0.36$) single crystals shows rather complex behavior. The values are notably smaller than that observed in  Ba(Fe$_{1-x}$Co$_x$)$_2$As$_2$ and are more consistent with those expected in normal, weakly correlated, metals. The $x - T$ phase diagram obtained from TEP measurements is similar to the previously outlined. \cite{tha10a} TEP analysis suggests two concentration ranges, $x \sim 0.07$ and $x \sim 0.3$ where Lifshitz transitions, or other significant changes of the electronic structure or correlations  might occur. Similar analysis of extended TEP data for  Ba(Fe$_{1-x}$Co$_x$)$_2$As$_2$ suggested, in addition to the known Lifshitz transition at $0.020 < x < 0.024$, two other concentration ranges, $x \sim 0.11$ and $x \sim 0.22$, where significant changes of the electronic structure or correlations possibly occur.  Detailed experimental (including ARPES and $c$-axis TEP) and theoretical studies in the vicinity of these critical concentrations are required to shed light on evolution of physical properties of BaFe$_2$As$_2$ with isoelectronic doping in comparison to the electron doping. These studies might be relevant for understanding of the results obtained under pressure as well. 

\begin{acknowledgments}
Ames Laboratory is operated for the U.S. Department of Energy by Iowa State University under Contract No. DE-AC02-07CH11358. This work was supported by the U.S. Department of Energy, Office of Basic Energy Science, Division of Materials Sciences and Engineering.  SLB and PCC were supported in part by the State of Iowa through the Iowa State University. Discussions (PCC, SLB) about "TEP as a wonder measurement" with Bryan Coles are fondly remembered. Help of A. Kracher and W. E. Straszheim in the elemental analysis of the crystals and of J. Q. Yan, N. Ni and S. Ran in synthesis  is greatly appreciated.
\end{acknowledgments}

\clearpage

\clearpage

\clearpage

\clearpage

\clearpage

\clearpage

\end{document}